\documentclass[useAMS,usenatbib]{mn2e}
\usepackage{graphicx}
\usepackage{url,mathtools}
\usepackage{lastpage}

\title[Disc properties and accretion rates of low-mass objects]
{The properties of discs around planets and  brown dwarfs as evidence for disc fragmentation}

\author[Stamatellos \& Herczeg]
{Dimitris Stamatellos$^{1,}$\thanks{E-mail:D.Stamatellos@astro.cf.ac.uk}, Gregory J. Herczeg$^{2}$ \\
$^1$ Jeremiah Horrocks Institute for Mathematics, Physics \& Astronomy, University of Central Lancashire, Preston, PR1 2HE, UK\\
$^2$\ Kavli Institute for Astronomy and Astrophysics, Peking University, Yi He Yuan Lu 5, Haidian District, Beijing 100871, China\\
 }

\begin{document}

\date{Accepted 2014 . Received 2014 July 14; in original form 2014 July 14}

\pagerange{\pageref{firstpage}--\pageref{lastpage}} \pubyear{201-}

\maketitle

\label{firstpage}

\begin{abstract}

  Direct imaging searches  have revealed many very low-mass objects, including a small number of planetary mass objects, as  wide-orbit companions  to young stars.  The formation mechanism of these objects remains uncertain.  In this paper we present the predictions of the disc fragmentation model regarding the properties of the discs around such low-mass objects. We find that the discs around objects that have formed by fragmentation in  discs hosted by Sun-like stars (referred to as {\it parent} discs and {\it parent} stars) are more massive than expected from the ${M}_{\rm disc}-M_*$ relation (which is derived for stars with masses $M_*>0.2~{\rm M}_{\sun}$). Accordingly, the accretion rates onto these objects are also higher than expected from the $\dot{M}_*-M_*$ relation. Moreover there is no significant correlation between the mass of the brown dwarf or planet with the mass of its disc nor with the accretion rate from the disc onto it.  The discs around objects that form by disc fragmentation have larger than expected masses as they  accrete gas from the disc of their parent star during the first few kyr after they form. The amount of gas that they accrete and therefore their mass depend on how they move in their parent disc and how they interact with it. Observations of  disc masses and accretion rates onto very low-mass objects are consistent with the predictions of the disc fragmentation model. Future observations (e.g. by ALMA) of disc masses and accretion rates onto substellar objects that have even lower masses (young planets and young, low-mass brown dwarfs), where  the scaling relations predicted by the disc fragmentation model 
 diverge significantly from the corresponding relations established for higher-mass stars, will test the predictions of this model.

\end{abstract}


\section{Introduction}

Many very low-mass objects, including a small number of planetary-mass objects, have been observed by direct imaging as companions to young stars at distances  from a few tens to a few hundred AU \citep{Kraus:2008a, Kraus:2013a, Marois:2008a, Faherty:2009a, Ireland:2011a,Kuzuhara:2011a,Kuzuhara:2013a, Aller:2013a,Bailey:2013a,Rameau:2013a,Naud:2014a,Galicher:2014a}. The dominant mechanism for the formation of low-mass stellar and substellar objects (low-mass hydrogen-burning stars, brown dwarfs and giant planets) is still uncertain \citep[e.g.][]{Chabrier:2014a,Stamatellos:2014a}. It is believed that such objects may form in three ways: (i) by collapsing molecular cloud cores, i.e. the same way as Sun-like stars \citep{Padoan:2004a,Hennebelle:2008c, Hennebelle:2009b, Hopkins:2013b}, (ii) by fragmentation of protostellar discs \citep{Boss:1997a, Stamatellos:2007c, Attwood:2009a, Stamatellos:2009a,Boley:2009a}, which may not even be centrifugally supported \citep{Offner:2010d, Offner:2012a}, and (iii) by ejection of proto-stellar embryos from their natal cloud cores \citep{Reipurth:2001a, Bate:2002a,Goodwin:2004a}. Additionally, gas giant planets also form by core accretion, i.e. by coagulation of dust particles to progressively larger bodies  \citep{Safronov:1969a,Goldreich:1973a,Mizuno:1980a,Bodenheimer:1986a,Pollack:1996a}. Objects formed by core accretion may even become deuterium-burning brown dwarfs \citep[e.g.][]{Molliere:2012a}. However, gas giants on wide orbits ($\stackrel{>}{_\sim}100-300~{\rm AU}$) are not believed to be able to form, at least in-situ, by core accretion.

Substellar objects are difficult to form similarly to Sun-like stars, and it has been argued that a different mechanism may in fact be at play \cite[e.g.][]{Whitworth:2007a, Thies:2007a,Reggiani:2013a}.
A low-mass pre-(sub)stellar core has to be very dense and compact in order to be gravitationally unstable. 
Up to now, only one clear-cut  self-gravitating brown dwarf-mass core has been observed \citep[][]{Andre:2012a}, but such cores have small size and they are faint, making them difficult to observe. Another way to reach the high densities that are required for the formation of substellar objects is in the discs around young stars. This model has been studied extensively and has been shown to reproduce critical observational constraints such as the low-mass IMF, the brown dwarf desert, and the binary statistics of low-mass objects \citep{Stamatellos:2009a, Lomax:2014a, Lomax:2014b}. In the third formation scenario mentioned in previous paragraph, formation by ejection of proto-stellar embryos, objects that were destined to become Sun-like stars fail to fulfil their potential as they are ejected from their natal cloud before they accrete enough mass to become hydrogen-burning stars.

The presence of discs around substellar objects (and associated phenomena, i.e. accretion and outflows) was initially thought to favour a Sun-like formation mechanism (i.e. turbulent fragmentation and collapse of pre-substellar cores). However, all three main formation mechanisms produce substellar objects that are surrounded by discs, albeit with different disc fractions. In the turbulent fragmentation scenario substellar objects almost always form with discs \citep[e.g.][]{Machida:2009a}. Substellar objects that form by disc fragmentation also most likely form with discs but these discs may be disrupted as these objects are liberated from the disc in which they formed \citep{Stamatellos:2009a}. In the ejection scenario discs are also likely to be disrupted but quite a few  still survive. \cite{Bate:2009b} finds that at least 10\% of the very-low mass objects formed in his simulations have discs with sizes larger $>40$~AU.

Although the presence of discs around substellar objects is consistent with all three formation theories, the properties of these discs may hide clues regarding their formation mechanism.  Recently, many authors \citep{Andrews:2013a, Mohanty:2013a, Ricci:2014a, Kraus:2014a} have estimated the masses of discs 
around young low-mass stellar and substellar objects down to a limit of $\sim 10^{-3}~{\rm M}_{\sun}$, using submillimetre observations . The accretion rates around many low-mass objects have also been determined down to  $10^{-13}$~M$_{\sun}\ {\rm yr}^{-1}$ \citep{Natta:2004a, Calvet:2004a, Mohanty:2005a, Muzerolle:2005a, Herczeg:2008a, Antoniucci:2011a,Rigliaco:2011a,Biazzo:2012a}. 

The goal of this paper is to compare these observations with the theoretical predictions of the disc fragmentation model.   This is particularly topical as the discovery  of many planetary-mass objects at wide separations (a few tens to a few hundred AU) from their host stars by direct imaging \citep{Kraus:2008a, Kraus:2013a, Marois:2008a, Faherty:2009a, Ireland:2011a,Kuzuhara:2011a,Kuzuhara:2013a, Aller:2013a,Bailey:2013a,Rameau:2013a,Naud:2014a,Galicher:2014a} has renewed the debate whether these objects have formed by core accretion or by fragmentation in the discs of their parent stars, or have formed otherwise and were later captured by the parent stars \citep{Perets:2012a}.  It is  also uncertain whether such companions may have formed differently than field objects. More wide-orbit substellar objects are bound to be discovered with focused surveys looking for giant planets  (Gemini Planet Imager, \citeauthor{Macintosh:2014a} 2014; SPHERE/VLT, \citeauthor{Beuzit:2008a} 2008; HiCIAO/SUBARU, \citeauthor{Suzuki:2009a} 2009) and therefore their properties and the properties of their probable discs may be better determined in the near future, providing tighter constraints for theoretical models.

In this paper we present the predictions of the disc fragmentation model regarding the masses of discs around low-mass stellar and substellar objects (brown dwarfs and planets) that are either companion to higher mass stars or free-floating. We also determine the accretion rates onto low-mass objects and compare them with observations.  In Section~\ref{sec:model} we briefly review the hydrodynamic simulations that we use for this study, and in Section~\ref{disc:evolution} we discuss how we compute the evolution of the discs around brown dwarfs and planets, after these discs have separated from the discs of their parent stars.  In Section~\ref{sec:discmass} we present the results of the model regarding the disc masses of low-mass objects and discuss how they fit with observations, and in Section~\ref{sec:accretion} we discuss the accretion rates onto low-mass objects. Finally in Section~\ref{sec:conclusions} we summerize the main results of this work.

\section{Simulations of the formation of  wide-orbit planets and brown dwarfs by disc fragmentation}
\label{sec:model}

\subsection{Overview}
The properties of the low-mass stellar and sub-stellar objects (planets, brown dwarfs, and low-mass hydrogen burning stars) formed by disc fragmentation have been studied  in detail by Stamatellos et al.  in a series of papers \citep{Stamatellos:2007c, Stamatellos:2009a,Stamatellos:2009d,Stamatellos:2011a}. In this paper we use the results of the simulations of \cite{Stamatellos:2009a} to determine the properties of the discs around wide-orbit planets, brown-dwarfs and low-mass stars that form in the discs of Sun-like stars, and the accretion rates onto these objects.

\subsection{Initial Conditions}
\cite{Stamatellos:2009a} performed 12 simulations of gravitationally unstable discs around Sun-like stars. These simulations are different realisations of the same star-disc system, i.e. the properties of the system are the same in all simulations; the only difference is the random seed used to construct each disc.  The star has an initial mass of $M_*=0.7$~M$_{\sun}$. The disc around it has an initial mass of $M_{\rm D}=0.7$~M$_{\sun}$ and a radius of $R_{\rm D}=400~$AU. The surface density of the disc is 
\begin{equation}
\Sigma_{_0}(R)=\frac{0.014\,{\rm M}_{\sun}}{{\rm AU}^2}\,\left(\frac{R}{\rm AU}\right)^{-7/4}\,,
\end{equation}
and its temperature
\begin{equation}\label{EQN:TBG}
T_{_0}(R)=250\,{\rm K}\,\left(\frac{R}{\rm AU}\right)^{-1/2}+10~{\rm  K}\,.
\end{equation}
The disc has an initial Toomre parameter $Q\sim 0.9$  and  therefore it is gravitationally unstable by construction. In a realistic situation the disc forms around a young protostar and grows in mass by accreting infalling material from the envelope \cite[e.g][]{Attwood:2009a, Stamatellos:2011e, Stamatellos:2012a, Lomax:2014a}. The disc fragments once it has grown enough to  become gravitationally unstable at distance $\sim 100$~AU from its parent star and this happen before it can reach the mass assumed by \cite{Stamatellos:2009a}.  In fact even discs with masses $\sim0.25$~M$_{\sun}$ and radii $100~$AU can fragment  \citep{Stamatellos:2011d}. Such disc masses are comparable to the observed disc masses in young (Class 0, Class I) objects \citep[e.g.][]{Jorgensen:2009a,Tobin:2012b, Murillo:2013a, Favre:2014a}. In any case, any evolutionary period with such a massive disc is short-lived as  the disc quickly (within a few thousand years) fragments. 

The large disc mass and size assumed by \cite{Stamatellos:2009a} ensure that more low-mass objects form in the disc to improve the statistical analysis of the results, but the properties of these objects (mass, disc mass, disc size) are similar to the ones formed in lower mass discs \citep{Stamatellos:2011d}. This is because the characteristic initial mass of  objects formed by disc fragmentaton  is set by  the opacity limit, which is thought to be $\sim1-5~{\rm M_{\rm J}}$ \citep{Low:1976a,Rees:1976a,Silk:1977a,Boss:1988a,Boyd:2005a,Whitworth:2006a,Boley:2010b,Kratter:2010b,Forgan:2011b,Rogers:2012a}. Therefore, the typical initial mass of the objects formed by fragmentation is the same for lower and higher parent disc masses. The parent disc mass (in lower mass discs)  is distributed among fewer objects and therefore the masses of these objects and the masses of their discs are similar to the ones that form  in  higher mass discs. 

 The simulations that we use start off with already formed discs; therefore disc loading and other interactions with the star forming cloud (which may lead to non-axisymmetric discs) are ignored. Simulations that  take these effects into account \citep[e.g.][]{Tsukamoto:2015a}
have given similar results to the simulations of \cite{Stamatellos:2009a} used in the present paper. We therefore do not anticipate the choice of the specific set of disc simulations to significantly alter the main conclusions of this paper. 

\subsection{Numerical Method}
The evolution and fragmentation of the disc of the parent star is followed using the SPH code {\sc dragon} which treats the radiation transport within the disc with the  diffusion approximation of \cite{Stamatellos:2007b}\citep[see also][]{Forgan:2009b}. The radiation feedback from the parent star is also taken into account.
The code uses time-dependent viscosity with parameters $\alpha=0.1$, $\beta=2\alpha$  (Morris \& Monaghan 1997) and a Balsara switch (Balsara 1995). 

\subsection{Results}

The parent disc is unstable and therefore within a few kyr it fragments into 5-11 secondary objects.  In the 12 simulations  a total of 96 objects are formed.  Some of them escape and others remain bound to the parent star at wide orbits \citep[see][]{Stamatellos:2009a,Stamatellos:2011a}. Most of these objects are brown dwarfs (67\%; $13~{\rm M}_{\rm J}< M<80~{\rm M}_{\rm J}$) and the rest are low-mass hydrogen burning stars (30\%; $M>80~{\rm M}_{\rm J}$), and planets (3\%; $M<13~{\rm M}_{\rm J}$).  These mass ranges are set by the hydrogen-burning limit ($\sim 80~{\rm M}_{\rm J}$), and the deuterium-burning limit ($\sim 13~{\rm M}_{\rm J}$). Stars can sustain hydrogen burning, whereas brown dwarfs can sustain only deuterium burning.  Planets cannot sustain deuterium burning. However, there is  no reason for gas fragmentation to stop either at the hydrogen-burning limit or the deuterium-burning limit: the minimum mass of an object that forms by gas fragmentation is given from the opacity limit for fragmentation ($\sim1-5~{\rm M_{\rm J}}$). On the other hand, planets that form by core accretion may have masses $>13~{\rm M}_{\rm J}$ \citep[e.g.][]{Molliere:2012a}. In this paper, we use the  term {\it planet} to refer  to objects with mass  $<13~{\rm M}_{\rm J}$  regardless on their formation mechanism.

\begin{figure}
\centerline{
\includegraphics[height=1.1\columnwidth,angle=-90]{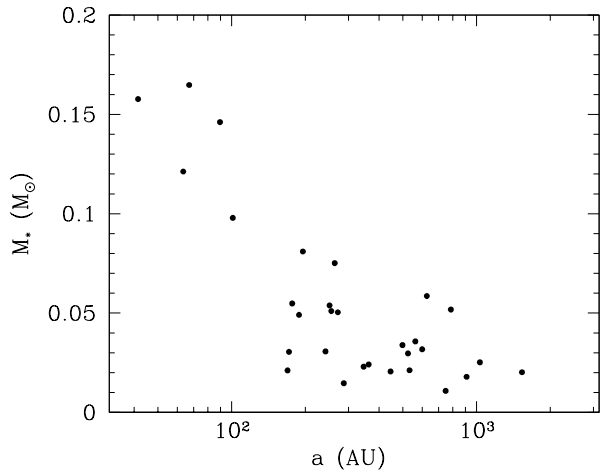}}
\caption{The masses of objects formed by disc fragmentation plotted against the semi-major axes of their orbits around the parent star. Most of these objects are brown dwarfs (with a few of them near the brown dwarf-planet boundary of 13~${\rm M}_{\rm J}$) and a few of them are planets and low-mass hydrogen-burning stars. Low-mass hydrogen burning stars tend to be closer to the parent star than brown dwarfs and planets.}
\label{fig:starmass.r}
\end{figure}

About 70\% of the secondary objects that form in the parent disc are attended by their own individual discs. These discs have masses up to a few tens of ${\rm M}_{\rm J}$ and radii of a few tens of AU \citep[see][]{Stamatellos:2009a}. Out of these secondary objects with discs we select 34 single objects (i.e. they are not in a binary system with another secondary object formed in the parent disc but they may still be bound to the parent star) for which the properties of the discs can be determined (i.e. the discs are nearly Keplerian). The rest of the objects either  were binaries, or were attended by disc-like structures whose properties could not be obtained (e.g. discs that were perturbed). Almost all of the objects in the sample  (33 out of 34) are still bound to the parent star albeit in most cases at very wide-orbits (see Fig.~\ref{fig:starmass.r}). Eventually many of these will  be liberated and will become field objects \citep{Stamatellos:2009a}. Therefore, in Sections \ref{sec:discmass} \& \ref{sec:accretion}, the properties of the discs of these objects and the accretion rates onto them will be compared with the observed properties of objects that are either wide-orbit companions to other stars, or field objects.



 Fig.~\ref{fig:starmass.r} presents the relation between the masses and  the semi-major axes of the orbits of these objects. Most of these objects are brown dwarfs (with a few of them near the brown dwarf-planet boundary of 13~${\rm M}_{\rm J}$) and a few of them planets and low-mass hydrogen-burning stars. Low-mass hydrogen burning stars tend to be closer to their parent stars than brown dwarfs and planets \citep[the brown dwarf desert;][]{Marcy:2000a, Grether:2006a, Sahlmann:2011a, Ma:2014a}. There are many brown dwarf  companions to Sun-like stars but at these tend to be at wide separations \citep{Kraus:2008a, Faherty:2009a, Faherty:2010a, Kraus:2011a, Evans:2012a,Reggiani:2013a,Duchene:2013b}.  As the above types of objects all form by the same mechanism in the disc irrespective of their mass we will analyse their disc properties collectively. 
 
\begin{figure}
\centerline{
\includegraphics[height=1.1\columnwidth,angle=-90]{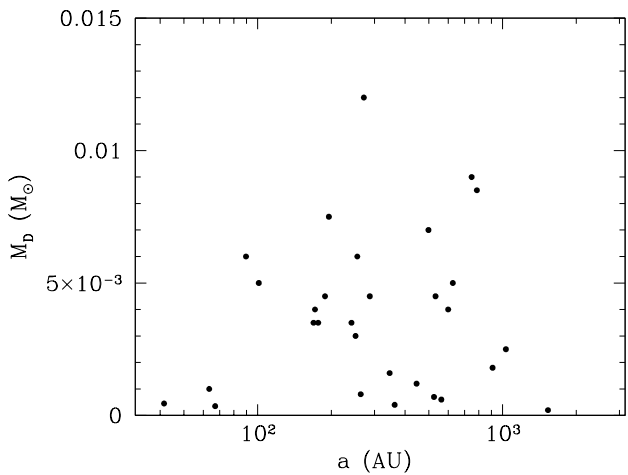}}
\caption{The disc masses around objects formed by disc fragmentation against the semi-major axes of their orbits around the parent star. There is no significant correlation between the two. The disc masses are probably determined by how these objects move in the disc of the parent star and accrete mass from it, rather than where they form in the parent disc.}
\label{fig:discmass.r}
\end{figure}

 Fig.~\ref{fig:discmass.r} presents the mass of each disc versus the semi-major axis of its host object. There is no significant correlation between  the two. The disc masses are determined by how these objects move within the disc of the parent star and  accrete mass from it, rather than where they form in the parent disc.
 
\section{The evolution of the discs around brown dwarfs and planets}
\label{disc:evolution}

The hydrodynamic simulations provide the  properties of the discs around wide-orbit companions to Sun-like stars at the time when 70-80\% of the parent disc around the parent Sun-like star has been accreted, either onto the parent star or onto the low-mass objects that form in the parent disc.  This typically happens within $10-20$~kyr from the start of each simulation. By this point the mass of the parent disc has been reduced to  $<0.01~{\rm M}_{\sun}$. Considering that the secondary objects that formed in the parent disc are on wide-orbits around the parent star, we do not expect interactions between the parent disc and the secondary discs to be important.  Additionally, in a cluster environment they are likely to be disrupted by stellar flybys  and become free-floating objects\cite[][]{Heggie:1975a,Kroupa:2003a,Parker:2009a, Parker:2009b, Spurzem:2009a, Malmberg:2011a,Hao:2013a}.
Therefore, we assume that at this point (i) that the secondary discs (i.e. the discs around the low-mass objects that form in the parent disc) have separated from their parent disc, (ii) that they evolve independently (i.e. there are no dynamical interactions between them and the parent disc, or other objects the form in the parent disc), and (iii) that no further mass from the parent disc is accreted onto them. These assumptions are not critical as the accretion of additional material onto the secondary disc reinforces our conclusions. To compare the properties of the discs around these low-mass companions with the observed disc properties of companions  in nearby young stellar clusters (age $\sim 1-15$~Myr) these properties need to be evolved in time. As this is not possible to be done by hydrodynamic simulations due to the large computational cost, we have employed an analytic model of viscous disc evolution. 

We  ignore any disc clearing due to photo-evaporation from radiation from the low-mass object hosting the disc \citep[see][and references therein]{Alexander:2013a}.  Photo-evaporation of  discs around low-mass objects ($\stackrel{<}{_\sim}0.15$~M$_{\sun}$) could happen \citep[e.g.][]{Alexander:2006a} but because of our limited knowledge on how UV and X-ray emission from low-mass objects would affect their discs, it is difficult to ascertain how important photo evaporation is for disc dispersion.

The analytic model we employ assumes  the disc (around a secondary object) is geometrically thin and evolves viscously under the influence of the central object's gravity \citep[e.g.][]{Lynden-Bell:1974a}, which in this case is the planet or brown dwarf  (represented as point masses in the model). The surface density of such a disc $\Sigma(R,t)$ at  polar radius $R$ and time $t$, evolves as follows
\begin{equation}
\frac{\partial\Sigma}{\partial t}=\frac{3}{R}\frac{\partial}{\partial R}\left[
R^{1/2}\frac{\partial}{\partial R}(\nu\Sigma R^{1/2})\right]\,,
\end{equation}
where  $\nu(R,t)$ is the kinematic viscosity \citep{Pringle:1981a}.  In this equation (and in subsequent equations) $t=0$ corresponds to the time where these discs are decoupled from their parents discs  (i.e. the end of the hydrodynamic simulations).
Assuming that the viscosity is independent of time and can be expressed as a power law in $R$, $\nu\propto R^\gamma$, then the above evolution equation has a similarity solution \citep{Lynden-Bell:1974a,Hartmann:1998a}
\begin{equation}
\Sigma(R,t)=\frac{M_{\rm d}(0)(2-\gamma)}{2\pi R^2_0 r^\gamma}\tau^{(5/2-\gamma)/(2-\gamma)} exp\left[-\frac{r^{2-\gamma}}{\tau}\right]\,, 
\end{equation}
where $r=R/R_0$ ($R_0$ is the radius within which 60\% of the disc mass is contained initially), and 
\begin{equation}
\label{eq:time}
\tau=t/t_\nu+1 \,,\end{equation}
where
\begin{equation}
\label{eq:tvisc}
t_\nu=R_0^2/ [3(2-\gamma)^2\nu(R_0)]\, .
\end{equation}
The accretion rate onto the central object is then
\begin{equation}
\label{eq:accretion}
\dot{M}_*=\frac{M_{\rm d}(0)}{2(2-\gamma) t_\nu}\tau^{-(5/2-\gamma)/(2-\gamma)}\,,
\end{equation}
and the disc mass
\begin{equation}
\label{eq:dmass}
M_{\rm d}(t)=M_{\rm d}(0) \tau^{-1/[2(2-\gamma)]}\,.
\end{equation}

It has been argued that observations of the discs of T~Tauri stars suggest that $\gamma\sim1$ \citep{Hartmann:1998a} (i.e. $\nu\propto R$), and therefore we will adopt this value in the present study. The choice of $\gamma$ is not critical for the conclusions of this paper.

 We use the $\alpha$-viscosity parametarisation \citep{Shakura:1973a}
\begin{equation}
\label{eq:visc}
\nu=\alpha c_{\rm s} H\,,
 \end{equation}
 where $c_{\rm s}$ is the sound speed in the disc, $H$ is the disc scale-height, and $\alpha$ the viscosity parameter. Assuming that the disc is locally vertically isothermal we obtain 
$
 H={c_{\rm s}}/{\Omega(R)},
$
 which when used in  Eq.~(\ref{eq:visc}) and assuming Keplerian rotation, i.e. $\Omega(R)=(GM_*/R^3)^{1/2}$, gives
 \begin{equation}
 \label{eq:visc2}
 \nu \propto \alpha\ T_d\ R^{3/2} M_*^{-1/2}\,.
 \end{equation}
 Using Eq.~(\ref{eq:visc2}) in Eq.~(\ref{eq:tvisc}) and assuming $T_d(R)\propto R^{-1/2}$ (consistent with  $\gamma=1$) we obtain
\begin{equation}
\label{eq:tvisc2}
\mathclap{t_\nu=8\! \times\! 10^4 \! \left(\frac{\alpha}{10^{-2}}\right)^{-1}\! \! \left(\frac{R_0}{10{\rm AU}}\right)\!\!  
\left(\frac{M_*}{0.5{\rm M}_{\sun}}\right)^{1/2}\!\! \!  \left(\frac{T_d}{10~{\rm K}}\right)^{-1} {\rm yr}} 
\end{equation}
where $T_d$ is the disc temperature at 100 AU.

We can therefore calculate the disc mass and the accretion rate onto the central object that hosts the disc (planet or brown dwarf) at any given time, using the initial disc mass $M_{\rm d}(0)$, obtained by the SPH simulations, and using Eqs.~(\ref{eq:dmass}),(\ref{eq:accretion}), and (\ref{eq:tvisc2}).

\section{The masses  of discs around low-mass stellar and substellar objects}
\label{sec:discmass}

Observations of disc masses \citep[e.g][]{Andrews:2013a, Mohanty:2013a} over a wide range of host stellar and substellar masses from intermediate mass stars to planetary-mass objects suggest  a linear  correlation between object mass and disc mass, i.e. $M_{\rm d}\propto M_*$.

\cite{Andrews:2013a} using 3 different evolutionary models for calculating stellar masses they find that stellar\footnote{In this context the terms {\it star} and {\it stellar} are used to refer to any objects formed by gravitational instability, therefore  including brown dwarfs and planets, as well as hydrogen-burning stars \citep{Whitworth:2007a}.}   mass scales almost linearly with the disc mass, $M_{\rm d}\approx 10^\kappa M_*^{\lambda}$, where $
\kappa=-2.3 \pm 0.3, -2.7 \pm 0.2,  -2.5 \pm 0.2$,  and
$\lambda=1.4\pm0.5, 1.0\pm0.4, 1.1\pm0.4$, when using the \cite{DAntona:1997a} (hereafter DM97), \cite{Baraffe:1998a} (BCAH98)  and \cite{Siess:2000a} (SDF00) models, respectively.
\cite{Mohanty:2013a} follow  a similar approach using the SDF00 models for stars with mass $>1.4~{\rm M}_{\sun}$, the BCAH98 model for stars with masses $0.08-1.4~{\rm M}_{\sun}$ and the dusty models of 
\cite{Chabrier:2000a} for stellar masses $<0.08~{\rm M}_{\sun}$, and similarly find that $M_{\rm d}\approx 10^{-2.4}M_*$. We note however that  both \cite{Andrews:2013a} and \cite{Mohanty:2013a} have assumed that the scatter in disc mass is constant for all objects irrespective of their mass; this may not be the case \citep{Alexander:2006c}.

It is evident (see Fig.~9 in \citeauthor{Andrews:2013a} 2013 and Fig.~9 in \citeauthor{Mohanty:2013a} 2013) that (i) there is a considerable scatter in the ${M}_{\rm disc}-M_*$ relation, and (ii) there are only a few definite detections of discs around stars with masses $<0.1~{\rm M}_{\sun}$. For example in the sample of \cite{Andrews:2013a} using the DCAH98 model there are just 15 definite disc detections around stars with mass $<0.1~{\rm M}_{\sun}$ (for 42 objects only upper limits for the disc masses were derived; these upper limits vary from  $6\times10^{-4}$ to $1.3\times10^{-2}~{\rm M}_{\sun}$).  
The large scatter in the data points and the small number of data points at low masses 
cast doubt to the suggestion that there is a  simple relation between stellar and disc mass that holds from intermediate-mass stars all the way to brown dwarfs and planetary-mass objects. In other words, these data do not exclude different scaling relations for low- and high-mass objects.
Another complication comes from the fact that when calculating the relation between stellar and disc masses the ages of these objects are not taken into account: disc masses are getting smaller with time either due to viscous evolution or due to photoevaporation from the host star. Thus, considering that discs around low-mass objects have masses that are low and near the detection limits of current observational facilities,  it is more likely to observe them when they are still young (and therefore have more mass).  Therefore, it may be expected that the discs around brown dwarfs and planets are more massive than what a simple extrapolation from the ${M}_{\rm disc}-M_*$  relation for higher mass stars would suggest.  The exact effect that the object ages have on the analyses of \cite{Andrews:2013a} and \cite{Mohanty:2013a} is difficult to estimate as stellar ages cannot be determined accurately enough \citep{Soderblom:2013a}.


\begin{figure}
\centerline{\includegraphics[height=1.15\columnwidth,angle=-90]{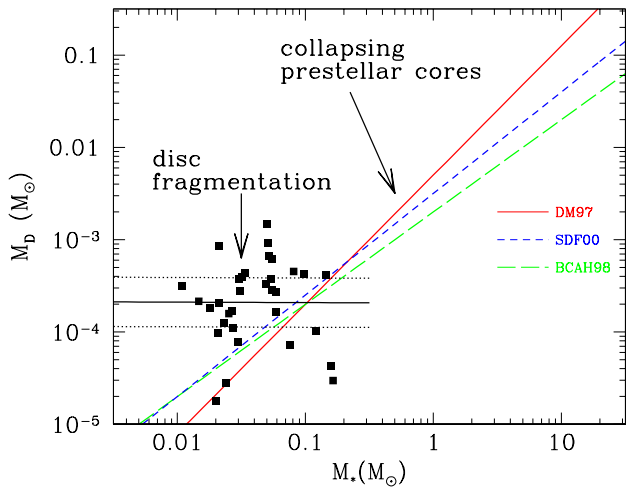}}
\caption{Disc masses of objects formed by disc fragmentation versus the masses of the host objects (black squares). The disc masses are calculated from the disc masses in the hydrodynamic simulations of Stamatellos et al. (2009a) assuming viscous disc evolution and  $\alpha=0.01$. The time for which each disc is evolved is chosen randomly between $1-10$~Myr, so as to emulate the age spread of observed discs. On the graph we also plot the best fit line (solid  black line) and the $\pm1\sigma$ region from the best fit (dotted black line).  The three coloured lines correspond the scaling relations derived by  Andrews et al. 2013 using different evolutionary models  (as marked on graph). The difference between the two relations indicate a different formation mechanism for low-mass objects. The difference is more pronounced at the extreme low-mass regime.}
\label{fig:sdmass}
\end{figure}

 The disc masses of the objects formed by disc fragmentation in the \cite{Stamatellos:2009a} simulations are plotted against the masses of the objects in Fig.~\ref{fig:sdmass}.  The disc masses are calculated from the disc masses in the hydrodynamic simulations of \cite{Stamatellos:2009a} assuming that the discs evolve viscously and using Eq.~(\ref{eq:dmass}) with $\alpha=0.01$.  The time for which each disc is evolved is chosen randomly between $1-10$~Myr, so as to emulate the age spread of  observed discs. The same figure shows the  relations  derived by \cite{Andrews:2013a} using  different evolutionary models to calculate the masses of the host star (DM97, BCAH98, and SDF00 as marked on graph).

There is scatter in the calculated disc masses of objects formed by disc fragmentation due to differences in the initial disc masses (i.e. the mass they have when they separate from the disc of the parent star) and their ages. Most of these discs are more massive than  expected from the scaling ${M}_{\rm disc}-M_*$ relation (which is mainly determined by higher mass stars) by more than a order of magnitude in a few cases. Additionally, there is no significant dependence between disc mass and stellar mass in contrast with higher-mass systems; we find a relation $\log (M_{\rm d})=-3.7-0.005\, \log(M_*)$ with a standard deviation of $\sigma=0.27$. 

\begin{figure}
\centerline{
\includegraphics[height=1.15\columnwidth,angle=-90]{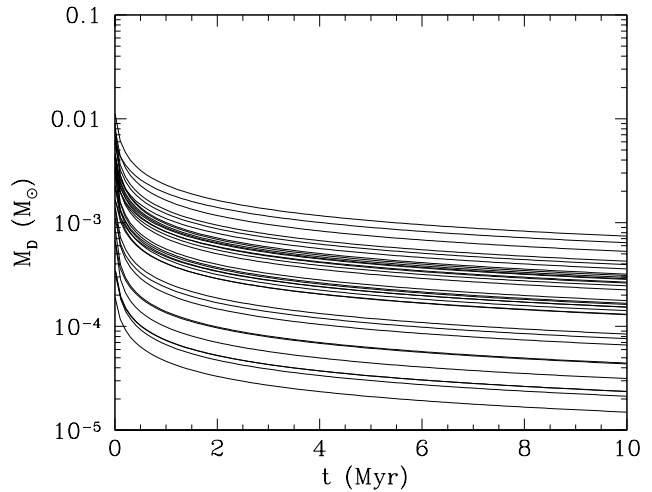}}
\caption{The evolution of the disc masses of the low-mass objects formed by disc fragmentation in the simulations of Stamatellos et al. (2009a). The disc masses are calculated assuming viscous disc evolution with a viscosity parameter $\alpha=0.01$. Disc mass decreases with time. Due to the difference in the initial disc masses there is a wide range of disc masses for each given time.}
\label{fig:dmasst}
\end{figure}

Both of the above characteristics are consequences of  formation by disc fragmentation. When a low-mass object forms from gas condensing out in the parent disc,  its properties (and its disc properties) are initially similar to an object that forms from a collapsing core in isolation. However as this object/disc system moves within the parent disc (but before it separates from the parent disc) it accretes more gas, and therefore its mass increases. This mass is initially accreted onto the object's disc and then slowly flows onto object. Therefore,  when a young object that has formed by disc fragmentation separates from its parent disc and evolves independently, has a  more massive disc than it would have if it had formed in isolation in a collapsing core. 

This scenario is consistent with the observations of \cite{Andrews:2013a} and \cite{Mohanty:2013a}; at least a few discs around young low-mass objects are more massive than expected. In their samples the detection limit is around $\sim 10^{-3}~{\rm M}_{\sun}$ and a few of the low-mass objects that they observed either have lower-mass discs or no discs at all. These may be  objects that have either formed by the collapse of a low-mass pre-(sub)stellar core like Sun-like stars, or objects that have formed by disc fragmentation but  have lost their discs (through evolution with time, see Fig.~\ref{fig:dmasst}, or due to interactions within the disc). The presence of low-mass discs around low-mass objects are consistent with both formation scenarios but the presence of relatively high-mass discs are indicative of formation by disc fragmentation.  Observations of disc masses around very low-mass objects ($\stackrel{<}{_\sim}10~{\rm M}_{\rm J}$), where the predicted ${M}_{\rm disc}-M_*$  relation  for young objects diverges significantly from the established ${M}_{\rm disc}-M_*$  relation derived for higher mass stars, will further test the disc fragmentation model.  ALMA has the required sensitivity and spatial resolution to observe such small discs. For example, \cite{Ricci:2014a},  have estimated disc masses down to $\sim 0.8-2.1~{\rm M}_{\rm J}$ in three young low-mass objects in the Taurus star forming region.

\begin{figure}
\centerline{
\includegraphics[height=1.15\columnwidth,angle=-90]{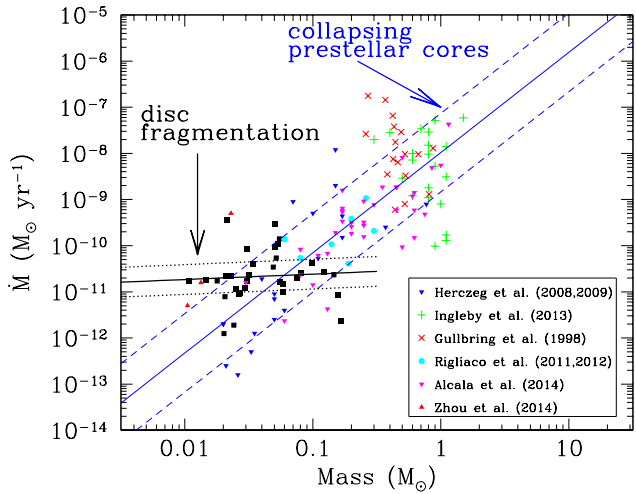}}
 \caption {The accretion rates  onto stars against their masses for a wide range of stellar masses. The black squares correspond to the objects formed by disc fragmentation in the simulations of Stamatellos et al. (2009) with accretion rates calculated assuming viscous disc evolution with $\alpha=0.01$.  The time for which each disc is evolved is chosen randomly between $1-10$~Myr, so as to emulate the age spread of observed discs. On the graph we also plot the best fit line (solid  black line) and the $\pm1\sigma$ region from the best fit (dotted black line). The remaining points  (coloured) correspond to observational data as marked on the graph. On the  graph we also plot the best fit line for the observations (solid  blue line) and the $\pm1\sigma$ region from the best fit (dashed blue lines) as estimated by Zhou et al. (2014).}
\label{fig:accretion}
\end{figure}

\section{Accretion rates onto wide-orbit low-mass objects}
\label{sec:accretion}

The accretion rates onto low- and higher-mass objects may also relate to their formation mechanism. In some cases it is possible to derive accretion rates even when the disc that provides the material for accretion is not detectable in the sub-mm, where disc masses are usually measured \citep{Herczeg:2009a,Joergens:2013a, Zhou:2014a}. For example, \cite{Herczeg:2009a} and \cite{Zhou:2014a} estimate the accretion luminosity from the excess line and continuum emission; for low-mass objects they can estimate accretion rates down to  $\sim10^{-13}$~M$_{\sun}\ {\rm yr}^{-1}$.

It has been argued that, similarly to the ${M}_{\rm disc}-M_*$ relation mentioned in the previous section, there is a relation between accretion rate onto a star and its mass. It has been suggested that this relation holds from intermediate-mass stars  down to brown dwarfs, namely that $\dot{M}_*\propto M_*^a$, where $\alpha\sim1.0-2.8$, albeit with a large scatter \citep{Natta:2004a, Calvet:2004a, Mohanty:2005a, Muzerolle:2005a, Herczeg:2008a, Antoniucci:2011a,Biazzo:2012a}.

\begin{figure}
\centerline{
\includegraphics[height=1.15\columnwidth,angle=-90]{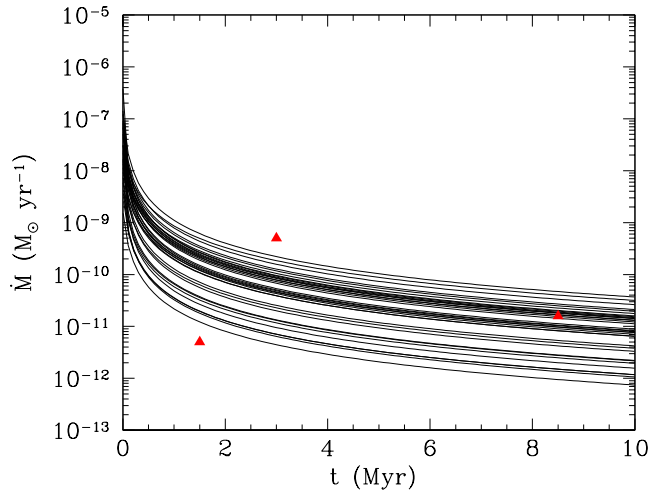}}
\caption{The evolution of the accretion rates of the objects formed by disc fragmentation in the simulations of Stamatellos et al. (2009a) . These accretion rates are calculated using the viscous evolution model  (Eq.~(\ref{eq:accretion}) with $\alpha=0.01$). There is a wide range of accretion rates for a specific age due to the spread in the initial disc masses. The three red triangles corresponds to the observations of Zhou et al. (2014). Considering the large uncertainties ($\sim 1-5$ Myr)  in the estimated ages these relatively high accretion rates are consistent with the predictions of  the disc fragmentation model.}
\label{fig:t-mdot}
\end{figure}

The accretion rates onto stars for a wide range of stellar masses are plotted against the stellar masses in Fig.~\ref{fig:accretion}. The accretion rates shown here have all been measured directly from excess Balmer continuum emission in the U-band \citep{Gullbring:1998a,Herczeg:2008a, Herczeg:2009a,Rigliaco:2011a, Rigliaco:2012a, Ingleby:2013a, Alcala:2014a, Zhou:2014a}. In the same figure the best fit line that was calculated by \cite{Zhou:2014a} is also plotted. It is evident from the graph that there is considerable scatter in $\dot{M}_*-M_*$ relation, that may reflect a difference in the disc initial conditions \citep{Alexander:2006c,Dullemond:2006a}. A part of the scatter  could also be attributed to the different ages of the systems plotted in  Fig.~\ref{fig:accretion}; accretion rates drop as stars age (see Fig.~\ref{fig:t-mdot}).

The detection limits of accretion rates are relatively low for planets and brown dwarfs. Most objects with excess  emission in the IR also have measured U-band accretion rates; thus it is expected that there is no bias towards detecting only younger objects  with higher accretion rates. In fact most of the observed objects exhibit low accretion rates. The estimated accretion rates for most of the low-mass objects  ($<0.1~{\rm M}_{\sun}$)  are consistent with the  $\dot{M}_*-M_*$ scaling relation derived for higher-mass stars. In fact in a few cases the accretion rates are lower than expected.  However in a few cases, like  the three planetary-mass companions observed by \cite{Zhou:2014a} (GSC 06214-00210 b, GQ Lup b, and DH Tau b) the accretion rates are higher than expected; these accretion rates  are an order of magnitude higher than what is expected from the $\dot{M}_*-M_*$ relation.

In Fig.~\ref{fig:accretion} we also plot the accretion rates of the objects formed by disc fragmentation in the simulations of \cite{Stamatellos:2009a}. These accretion rates are calculated using the viscous evolution model  (Eq.~(\ref{eq:accretion})) with $\alpha=0.01$. The time for which each disc is evolved is chosen randomly between $1-10$~Myr, so as to emulate the age spread of observed discs. There is no significant correlation  between the accretion rate and the mass of the object; we find a relation $\log (\dot{M}_*)=-10.5-0.12\, \log(M_*)$, with a standard deviation of $\sigma=0.3$. Moreover, in a few cases the accretion rates are higher than expected from the  $\dot{M}_*-M_*$ scaling relation. In the model that we present here, this is due to the  higher initial mass of the discs of these objects. As mentioned in the previous section,  these secondary discs grow in mass as they move within the discs of their parent stars (before they start evolving independently).  Therefore, we suggest that the relatively high accretion rates are indicative of formation by disc fragmentation. On the other hand,  low accretion rates are consistent with both formation by disc fragmentation or formation by the collapse of low-mass pre-(sub)stellar cores. In the former case low accretion rates could be due to time evolution (accretion rate drops with time; see Fig.~\ref{fig:t-mdot}) or due to disruption by interactions with other objects in the parent disc.

Observations of accretion rates around very low-mass objects \citep[$\stackrel{<}{_\sim}~10~{\rm M}_{\rm J}$; e.g.][]{Zhou:2014a}, where the predicted $\dot{M}_*-M_*$ relation relation diverges significantly from the established $\dot{M}_*-M_*$ relation  relation derived for higher mass stars, will further test the model presented here.

\section{The effect of the viscosity of secondary discs}

We have so far assumed in our analysis that the physical processes for redistributing angular momentum are the same for discs of T Tauri stars and for discs of lower mass objects (brown dwarfs, planets). However, this may not be the case. It has been argued that  the magneto-rotational instability may not be effective in discs around low-mass objects \citep{Keith:2014b,Szulagyi:2014a,Fujii:2014a}, which means that the effective viscosity in such discs should be smaller than the one presumed for T Tauri star discs ($\alpha=0.01$). However, these studies have focused on discs around Jovian planets on Jovian orbits, i.e. orbits relatively close to the central stars \citep[e.g.][]{Gressel:2013b}. In our study we focus on wide-orbit low-mass companions  (see Fig.~\ref{fig:discmass.r}), whose discs are more extended as they not limited by the Hill radii of their host secondary objects \citep[see][]{Stamatellos:2009a}. These discs could be massive enough so that angular momentum can be effectively transported by gravitational torques.

\begin{figure}
\centerline{
\includegraphics[height=1.15\columnwidth,angle=-90]{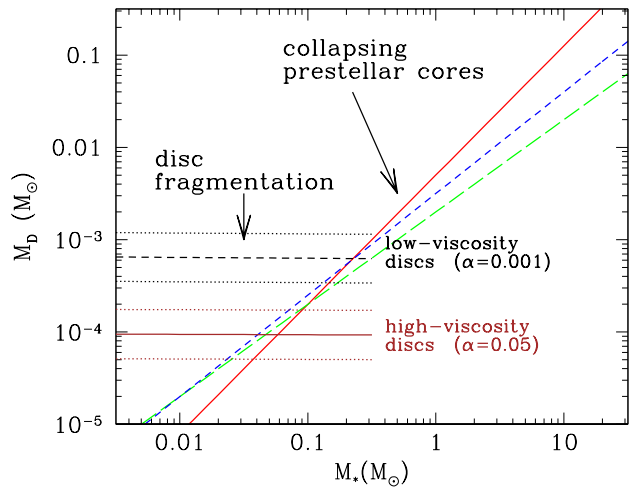}}
 \caption {Disc masses of objects formed by disc fragmentation versus the masses of the host objects for different disc viscosities. The best fit lines are calculated similarly to the ones in Fig.~\ref{fig:dmasst}, assuming that $\alpha=0.001$ (black solid line) or $\alpha=0.05$ (brown solid line). The dotted lines correspond to the $\pm1\sigma$ region from the best fit for each case.  The other three coloured lines correspond to the scaling relations derived by  Andrews et al. 2013 (see Fig.~\ref{fig:dmasst}). The differences between this relation and the ones derived in this paper (i.e. for objects formed by disc fragmentation) are more pronounced for low-viscosity secondary discs.}
\label{fig:vdmass}
\end{figure}

\begin{figure}
\centerline{
\includegraphics[height=1.15\columnwidth,angle=-90]{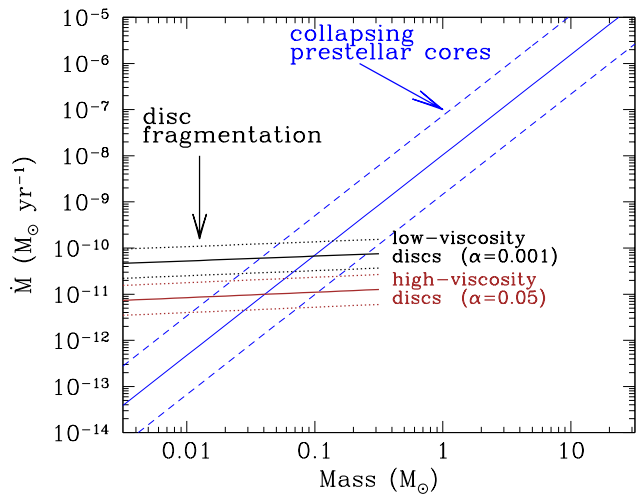}}
 \caption {The accretion rates  onto stars against their masses for a wide range of stellar masses. The best fit lines are calculated similarly to the ones in Fig.~\ref{fig:accretion}, assuming that $\alpha=0.001$ (black solid line) or $\alpha=0.05$ (brown solid line). The dotted lines correspond  to the $\pm1\sigma$ region from the best fit for each case.  On the  graph we also plot the best fit line for the observations (solid  blue line) and the $\pm1\sigma$ region from the best fit (dashed blue lines) (Zhou et al. 2014). The differences between this relation and the ones derived in this paper (i.e. for objects formed by disc fragmentation) are more pronounced for low-viscosity secondary discs.}
\label{fig:vaccretion}
\end{figure}

Nevertheless, our knowledge of the effective viscosity in such discs is limited, and it is important to examine the effect that the assumed disc viscosity has on the conclusions of our study. In Figs.~\ref{fig:vdmass} and \ref{fig:vaccretion} we present the predictions of our model for low-viscosity discs ($\alpha=0.001$) and for high-viscosity discs ($\alpha=0.05$). As expected, low-viscosity discs evolve slower and their masses and accretion rates remain higher for longer. Therefore, in this case the differences between the predicted  ${M}_{\rm disc}-M_*$ and $\dot{M}_*-M_*$ relations for disc fragmentation and the observed relations for higher mass stars are more pronounced (see black lines in Figs.~\ref{fig:vdmass}, \ref{fig:vaccretion}). The opposite holds for high-viscosity discs ($\alpha=0.05$; see brown lines in Figs.~\ref{fig:vdmass}, \ref{fig:vaccretion}).


\section{Conclusions}
\label{sec:conclusions}

 We suggest that substellar (planetary-mass objects and brown dwarfs)  and low-mass stellar objects  (low-mass hydrogen burning stars) that form by disc fragmentation, have disc masses and accretion rates that (i) are independent of the mass of the host object, and (ii) are  higher than what is expected from scaling relations derived from their intermediate and higher-mass counterparts. These low-mass objects form similarly to higher-mass objects by self-gravitating gas but as they move within the gas-rich parent disc their individual discs accrete additional material; therefore before these objects separate from their parent discs and evolve independently (i.e. within a few kyr), their discs grow more massive and the accretion rates onto them are higher than if they were formed in isolation in collapsing low-mass pre-(sub)stellar cores. The assumption of independent evolution is not critical as if these secondary discs were still interacting with their parent disc they would  accrete additional material reinforcing the above conclusion. However, we do not expect additional accretion to be important.

Observations of disc masses and accretion rates of low-mass objects  are consistent with the predictions of the disc fragmentation model. Although the presence of  low-mass discs (or lack of  discs) and low accretion rates (or no accretion at all) may be attributed to disc evolution and/or disc disruption due to interactions with other objects within the parent  disc, relatively high disc masses and high accretion rates are suggestive of formation due to disc fragmentation. We therefore suggest that low-mass objects that have discs with masses higher than expected  (or equivalently  accretion rates onto them higher than expected), such as  GSC 06214-00210 b, GQ Lup b, and DH Tau b \citep{Zhou:2014a}, are young objects that have formed by disc fragmentation.

The disc fragmentation model can further be tested by observations of disc masses and accretion rates of very low-mass objects ($\stackrel{<}{_\sim}10{\rm M}_{\rm J}$). At these very low-masses the 
${M}_{\rm disc}-M_*$  and $\dot{M}_*-M_*$ relations predicted by the model presented here 
 diverge significantly from the corresponding relations established for higher-mass stars. We suggest that future analyses of the ${M}_{\rm disc}-M_*$ and $\dot{M}_*-M_*$ relations should separate the sample into two subgroups, low-mass ($<0.2~{\rm M}_{\sun}$) and higher-mass ($>0.2~{\rm M}_{\sun}$) objects, so as to test whether these objects obey different scaling relations. 
 
 The intense interest in  wide-orbit  and free-floating planets has given momentum to the development of instruments with high sensitivity and good spacial resolution. Therefore observations in the near future are expected to deliver many more such low-mass objects. ALMA is already delivering such observations \citep{Ricci:2014a, Kraus:2014a}. The study of these objects, their disc properties and  the accretion rates onto them (if they are still young) will provide further constraints regarding their formation mechanism.

\section*{Acknowledgements}
We would like to thank the referee for many insightful comments that helped to improve the paper and clarify our main conclusions. We thank Yifan Zhou for important input in the paper and Richard Alexander for useful comments on photo-evaporation.  DS thanks Thijs Kouwenhoven for his hospitality during a visit to the  Kavli Institute for Astronomy \& Astrophysics at Peking University, where part of this work was completed.
\bibliography{../../bibliography}{}

\label{lastpage}

\end{document}